\documentclass[conference]{IEEEtran}   
\IEEEoverridecommandlockouts
\usepackage{cite}
\usepackage{amsmath,amssymb,amsfonts}
\usepackage{lipsum}
\usepackage{graphicx} 
  
\ifCLASSOPTIONcompsoc
\usepackage[caption=false,font=normalsize,labelfont=sf,textfont=sf]{subfig}
\else
\usepackage[caption=false,font=footnotesize]{subfig}
\fi
\usepackage{textcomp}
\usepackage{color}
\usepackage{xcolor}
\usepackage{array}
\usepackage{booktabs}
\usepackage{tabularx}

\usepackage{algorithm}  
\usepackage{algorithmicx}  
\usepackage{algpseudocode}

\def\BibTeX{{\rm B\kern-.05em{\sc i\kern-.025em b}\kern-.08em
		T\kern-.1667em\lower.7ex\hbox{E}\kern-.125emX}}

\usepackage{fancyhdr}
\usepackage{url}
\usepackage{soul}

\begin{document}

\title{String-Level Ground Fault Localization for TN-Earthed Three-Phase Photovoltaic Systems}


\author{\IEEEauthorblockN{
		Yuanliang Li\IEEEauthorrefmark{1},
		Xun Gong\IEEEauthorrefmark{1},
		Reza Iravani\IEEEauthorrefmark{2},
		Bo Cao\IEEEauthorrefmark{1},
		Heng Liu\IEEEauthorrefmark{1},
		Ziming Chen\IEEEauthorrefmark{1}}
\IEEEauthorrefmark{1}Huawei Montreal Research Centre, Huawei Technologies Canada Co., LTD, Montr\'eal, Canada\\
\IEEEauthorrefmark{2}Department of Electrical \& Computer Engineering, University of Toronto, Toronto, Canada
\thanks{This work has been submitted to the IEEE for possible publication. Copyright may be transferred without notice, after which this version may no longer be accessible. \par Corresponding Author: Xun Gong (xun.gong@huawei.com). }
}

\maketitle

\begin{abstract}
The DC-side ground fault (GF) poses significant risks to three-phase TN-earthed (3$\phi$-TN) photovoltaic (PV) systems, as the resulting high fault current can directly damage both PV inverters and PV modules. Once a fault occurs, locating the faulty string through manual string-by-string inspection is highly time-consuming and inefficient. 
This work presents a comprehensive analysis of GF characteristics through fault-current analysis and a simulation-based case study covering multiple fault locations. Building on these insights, we propose an edge-AI–based GF localization approach tailored for 3$\phi$-TN PV systems. A PLECS-based simulation model that incorporates PV hysteresis effects is developed to generate diverse GF scenarios, from which correlation-based features are extracted throughout the inverter’s four-stage shutdown sequence. Using the simulated dataset, a lightweight Variational Information Bottleneck (VIB)-based localization model is designed and trained, achieving over 93\% localization accuracy at typical sampling rates with low computational cost, demonstrating strong potential for deployment on resource-constrained PV inverters.
\end{abstract}

\section{Introduction}



\IEEEPARstart{T}{he} rapid growth of the photovoltaic (PV) industry has raised concerns about system reliability. Operating under harsh outdoor conditions, PV systems are prone to various faults (e.g., module degradation, shading, hot-spots, arcs, and ground faults), which can cause energy loss, component degradation, or even fires \cite{ieapvpsfailure2025}. 
Among these faults, \textbf{DC-side ground faults} (GFs), i.e., unintended connection of a conductor to earth in PV array, are particularly critical due to their severe impact and complex characteristics \cite{eskandari2020fault}. In a PV array, cables and module insulation will degrade over time and suffer damage from environmental stress, increasing the risk of a conductor contacting with grounded structures. Moreover, in multi-MPPT string inverter systems, where one inverter may connect to dozens of PV strings, locating GFs often requires inspecting each string manually, which is an expensive and time-consuming procedure. Thus, automated ground fault localization down to the string level is of great practical value \cite{mehmood2021fault}.

PV systems can be classified as grounded or ungrounded depending on whether one DC node is intentionally connected to the protective earth (PE) \cite{alam2015comprehensive}. Both types are widely used but behave differently under GFs. Grounded PV systems, typically used in single-phase residential setups, have one DC node (usually the negative terminal) tied to PE. When a GF occurs, the fault current flows through the DC-side grounding path, allowing quick detection and isolation by ground-fault protection devices (GFPDs), such as fuses \cite{flicker2013photovoltaic}. The maximum fault current is close to the PV module short-circuit current ($I_{\rm sc}$, generally $<$10 A), so inverter damage is limited. 

Several studies have been conducted on GFs in grounded PV systems. Ref.~\cite{flicker2016photovoltaic} revealed that fuse (or GFPDs)–based protection has a ``blind spot": GFs with high-impedance or low-current return paths may go undetected. Similarly, Ref.~\cite{pillai2019extended} argued that conventional protection standards (e.g., NEC) may be insufficient under mismatch or low-irradiance conditions, underscoring the need for more advanced detection techniques.
In \cite{alam2014pv} and \cite{roy2017irradiance}, spread-spectrum time-domain reflectometry (SSTDR) was used to detect GFs by detecting impedance discontinuities, which remains effective even under low or no irradiance and is independent of fault-current magnitude. Ref.~\cite{pillai2018mppt} leveraged the ``right-most power peak" feature of MPPT to detect GFs and demonstrated its effectiveness under mismatch, low irradiance, and partial shading.

\begin{figure}[hbtp] 
	\centering
	\includegraphics[width=1.0\columnwidth]{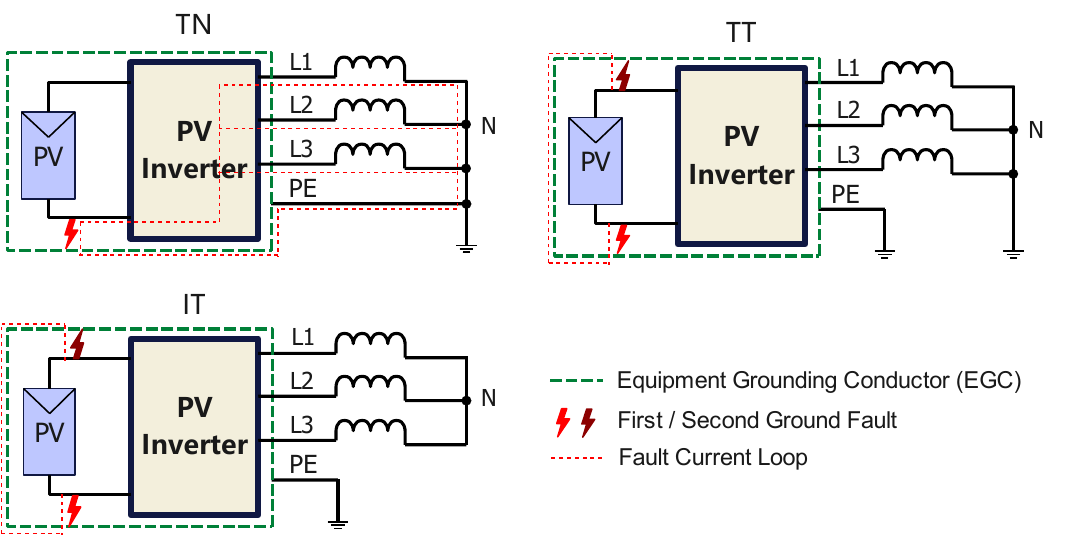}
	\caption{Ground faults in three-phase PV systems under three types of earthing arrangements, i.e., TN, TT, and IT.}
	\label{fig-earthing-types}
\end{figure}

Ungrounded PV systems, commonly used in three-phase commercial and industrial applications, have no DC node directly tied to PE. They can adopt three earthing arrangements: TN, TT, and IT \cite{earthingwiki}, as shown in Fig.~\ref{fig-earthing-types}.
In a TN system, the grid neutral (N) and all equipment grounding conductors (EGCs) in PV systems are earthed to PE, so a GF can cause a fault current to flow through the inverter, potentially damaging its internal circuitry. In a TT system, the grid neutral and EGCs of the PV system are separately earthed. In an IT system, the grid neutral is unearthed, and only the EGCs of the PV system are locally earthed. Both TT and IT systems can tolerate a first GF because no low-impedance loop forms between PE and N, though a second fault may still create a DC-side current loop with limited inverter impact. For IT or TT systems, Ref.~\cite{karmacharya2017fault} measures converter's midpoint voltage and applies the wavelet and artificial neural network (ANN) to locate GFs. 

Nevertheless, GF analysis and localization in three-phase TN-earthed (3$\phi$-TN) PV systems has received limited attention. This work targets this gap by developing a string-level GF localization method tailored for such systems and evaluating its feasibility for deployment on inverter microprocessors as an edge solution. Our \textbf{motivations} include:
\textbf{(1)} 3$\phi$-TN PV inverters are widely used in commercial and industrial settings \cite{vallve2016earthing}, where GFs constitute a notable portion of inverter failures.
\textbf{(2)} GFs in these systems can directly damage PV inverters and PV modules, causing safety risks and economic losses (see Section~\ref{section-system-model}).
\textbf{(3)} Modern multi-MPPT inverters connect many PV strings, making manual string-by-string GF inspection slow and costly. Under this scope, three main \textbf{challenges} arise:

\begin{itemize}
	\item \textbf{Transient and destructive nature.} In 3$\phi$-TN systems, a GF causes fault current to flow through the inverter almost instantaneously, which triggers the inverter's shutdown quickly. After that, offline diagnostics (such as I–V scans) cannot be performed due to potential internal failure, making high-frequency online sampling essential for analyzing transient fault data.
	
	\item \textbf{Complex fault characteristics.} \textbf{(1)} Fault waveforms vary substantially with GF locations, making conventional threshold-based methods unreliable. \textbf{(2)} GF impacts propagate across all strings, causing both faulty and healthy strings to deviate from nominal behavior, which complicates the discrimination. 
	
	\item \textbf{Data scarcity.} GF cases are rare in real world, and controlled GF experiments are destructive, making it difficult to obtain diverse, high-quality datasets for GF analysis and data-driven-based localization approaches.
	
\end{itemize}

To address these challenges, we propose an edge-AI–based approach for locating string-level GFs in 3$\phi$-TN PV systems. An offline edge-AI development pipeline with three phases is established.
In the first phase, a PLECS-based \cite{asadi2019simulation} simulation model tailored for 3$\phi$-TN PV systems is developed, where a dynamic PV model considering the hysteresis effect is adopted to enhance the transient representation of PV characteristics under GFs. This simulation model is used to generate diverse GF scenarios with various fault configurations, including fault locations, environmental parameters, fault triggering timestamps, etc. Furthermore, a four-stage shutdown procedure of PV inverter under GF is defined and modeled.
In the second phase, features extraction is performed on the simulated data. Correlations between the string current and other electrical signals are employed as the primary features. 
In the final phase, a lightweight Variational Information Bottleneck (VIB) model is developed as the edge-side classifier to determine the GF localization.

In brief, the \textbf{contributions} of this works are summarized as follows:
\begin{itemize}
	\item A dynamic PV model that accounts for the hysteresis effect of PV modules is applied to GF simulation for 3$\phi$-TN PV systems, thereby improving the modeling of PV transient behaviors under GFs.
	
	\item A comprehensive analysis of GF characteristics in 3$\phi$-TN PV systems is provided, which includes fault-current analysis in which lower-leg, upper-leg, and zero phase-fault-currents are defined and characterized, and a simulation-based case study for GFs at various locations.
	
	\item An offline edge-AI development pipeline for GF localization is proposed, comprising: (1) a data generation phase producing diverse GF cases under various fault configurations; (2) a feature extraction phase that leverages the correlations between string currents and other electrical variables; and (3) a lightweight VIB-based fault localization model. Finally, the trained model achieves over 93\% localization accuracy with low computational cost, demonstrating its feasibility for the deployment on resource-constrained edge devices.
\end{itemize}

The paper is organized as follows: Section \ref{section-system-model} presents the simulation model of the 3$\phi$-TN PV system under GFs where the fault currents will be analyzed and the dynamic PV model will be introduced. Section \ref{section-case_study} provides a case study on GFs at different fault locations. Section~\ref{section-ai} introduces the offline edge-AI development pipeline, including data generation, feature extraction, and the lightweight VIB-based classifier. Section~\ref{section-experiment} presents the experimental validation. Section~\ref{section-conclusion} concludes the paper and discusses future directions.


\begin{figure*}[t] 
	\centering
	\includegraphics[width=2.0\columnwidth]{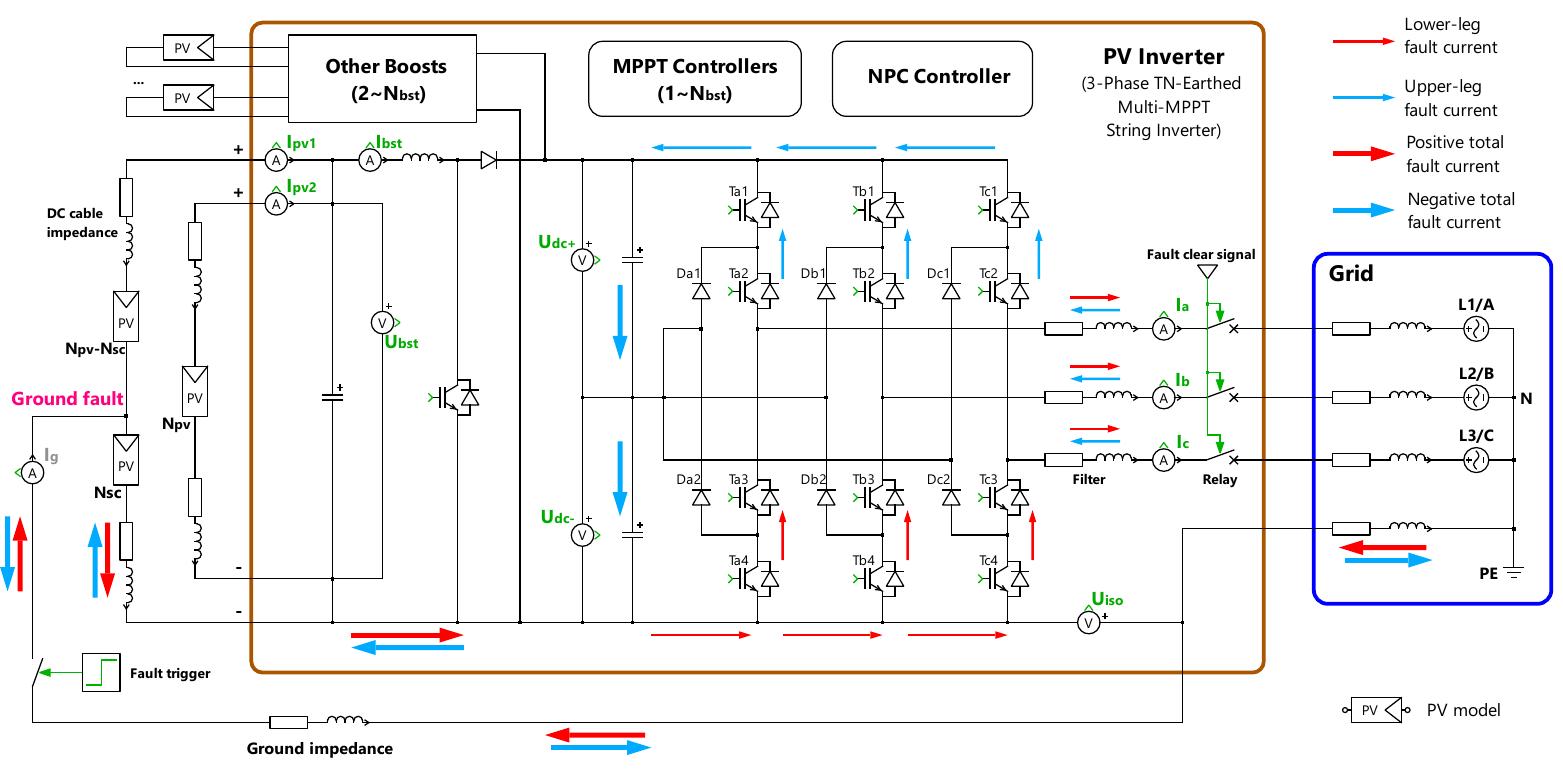}
	\caption{Diagram of the 3$\phi$-TN multi-MPPT PV inverter under GFs.}
	\vspace{-2mm}
	\label{fig-inverter_circuit}
\end{figure*}

\section{System Model and Fault-Current Analysis} \label{section-system-model}
Fig.~\ref{fig-inverter_circuit} presents the simulation diagram of the target 3$\phi$-TN PV system implemented in PLECS. Specifically, the studied PV inverter is a multi-MPPT string inverter with $N_{\rm bst}$ boost circuits.
Each boost circuit has multiple PV strings connected in parallel. Each boost circuit is controlled by a maximum power point tracking (MPPT) controller. The subsequent DC-AC circuit applies the 3-level I-type neutral point clamped (I-NPC) topology \cite{pouresmaeil2011control}, which has been increasingly adopted by many PV inverter manufacturers. The inverter is connected to the grid via harmonic filters and protective relays under a TN earthing arrangement. This DC-AC circuit is controlled by a NPC controller. Each PV string has $N_{\rm pv}$ PV modules connected in series, which are wrapped and simulated by a designed PV model. Moreover, DC cable impedance is also implemented at the positive and negative terminal of each string. 

For the GF simulation, we assume the GF occurs in one of the PV strings, potentially short-circuiting $N_{\rm sc} \in \{0,1,...,N_{\rm pv}\}$ PV modules within the string. $N_{\rm sc}=0$ indicates the fault is on the cable of the negative terminal. $N_{\rm sc}=N_{\rm pv}$ means the fault is on the cable of the positive terminal. The ground impedance is also modeled, representing the impedance between the fault point and the PE access point. The GF is initiated by using a switch with a configurable trigger time. 

The PV inverter is able to measure the current of each PV string ($I_{\rm pv}$), the inductor current of each boost ($I_{\rm bst}$), the input voltage of each boost ($U_{\rm bst}$), the DC link voltage ($U_{\rm dc} = U_{\rm dc+} + U_{\rm dc-}$), the insulation voltage ($U_{\rm iso}$, voltage between the negative terminal and PE), the output three-phase current ($I_{\rm a}$, $I_{\rm b}$, and $I_{\rm c}$), and the output three-phase voltage ($U_{\rm ab}$, $U_{\rm bc}$, and $U_{\rm ca}$).

\subsection{Fault-Current Analysis}
As depicted in Fig.~\ref{fig-inverter_circuit}, the 3$\phi$-TN PV system features the grounded neutral point on the grid side with all EGCs connected to this point. In the event of GF, fault currents can flow immediately through the inverter via the N-PE connection. Therefore, the fault current ($I_{\rm g}$) can be approximated as the sum of three phase currents:
\begin{equation}
	I_{\rm g} \approx I_{\rm a} + I_{\rm b} + I_{\rm c}
\end{equation}
A positive fault current ($I_{\rm g} > 0$) flows inward toward the fault location,  whereas a negative fault current ($I_{\rm g} < 0$) flows outward from it. The pattern of each phase's fault current is determined by the combined effect of the short-circuited PV modules and the corresponding grid phase voltage. The following fault-current analysis is carried out under the condition that all inverter IGBTs are in the non-conducting state.

\subsubsection{Lower-Leg Phase-Fault-Current}
For a certain phase, when the grid phase voltage ($U_{\rm g}$) plus the voltage provided by short-circuited PV modules ($U_{\rm sc}$) is below zero, a lower-leg fault current can be formed by a low-impedance path. This current will flow in reverse through the short-circuited PV modules, then pass forward through the freewheeling diodes of lower-leg IGBTs, subsequently flowing to the grid and returning to the fault point via the grounding system. 
A severe condition occurs when $U_{\rm g} < 0$ and few or no PV modules are short-circuited.

Under such conditions, the lower-leg IGBTs could be damaged within milliseconds by a large fault current. Additionally, the reverse overcurrent can also harm the PV cells, leading to either immediate failure or accelerated degradation \cite{sidawi2011effect}.

\subsubsection{Upper-Leg Phase-Fault-Current}
When the combined voltage $U_{\rm g} + U_{\rm sc}$ exceeds the DC-link voltage ($U_{\rm dc}$), the phase-fault-current will flow through the freewheeling diodes of the upper-leg IGBTs to charge the DC-link capacitors. It does no pass through the lower leg due to the diode block in lower-leg IGBTs. This situation typically occurs when a big portion of PV modules is short-circuited by the GF. 

Under such conditions, the DC-link capacitors may be at risk of overcharging, as they are energized simultaneously by the grid sources and short-circuited PV modules.

\subsubsection{Zero Phase-Fault-Current}
Phase-fault-current becomes zero when the combined voltage $U_{\rm g} + U_{\rm sc}$ is between zero and $U_{\rm dc}$. However, the total fault current $I_{\rm g}$ may remain nonzero since other phase voltages are different due to the phase-shift. 

A summary of the phase-fault-current patterns is presented in Table~\ref{tab:fault_current}. Therefore, GFs in 3$\phi$-TN PV systems can pose risks to the IGBTs, PV modules, and DC-link capacitors.

\renewcommand{\arraystretch}{1.2}
\begin{table}[htbp]
	\centering
	\caption{Summary of Phase-Fault-Current}
	\label{tab:fault_current}
	\begin{tabular}{cccc}
		\hline
		$U_{\rm g} + U_{\rm sc}$ & $<0$ & $\in [0, U_{\rm dc}]$ &  $>U_{\rm dc}$\\ 
		\hline
		\textbf{Phase-fault-current} & Lower-leg &  Zero & Upper-leg \\
		\hline
	\end{tabular}
\end{table}

\subsection{Dynamic PV Model} \label{subseciton-dynamic-pv}
\begin{figure}[h]
	\centering
	\includegraphics[width=0.68\columnwidth]{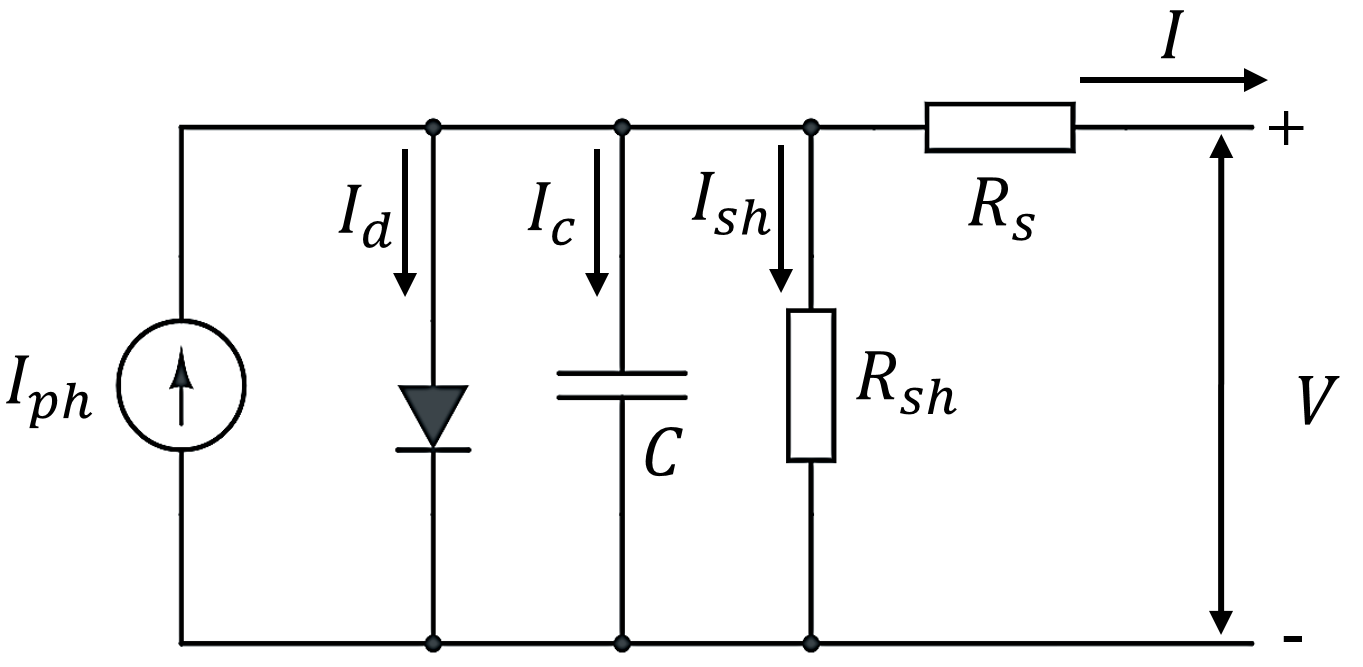}
	\caption{Dynamic PV cell model (single diode model with diffusion capacitor).}
	\label{fig-dsdm}
\end{figure}

Since PV modules are involved in the fault-current loop, an accurate simulation model of PV should be considered. However, traditional static single-diode model (SDM) \cite{jordehi2016parameter} does not consider the hysteresis effect caused by the diffusion capacitance of the P-N junction in PV cells. As a result, it cannot reflect the transient characteristics of PV modules under rapid current or voltage changes caused by GF events. 
To improve its transient representation for GF analysis, we adopt a dynamic PV cell model that can model the hysteresis effect by adding a diffusion capacitor to the SDM \cite{gao2018effects}. The equivalent circuit is shown in Fig.~\ref{fig-dsdm}.
\par
The I–V relationship of the model can be expressed as:
\begin{equation}\label{eq-dsdm}
		I  = 
		I_{\rm ph} 
		- \underbrace{ I_0 \left[ \exp \left( \frac{V+I R_{\rm s}}{nV_{\rm th}} \right) - 1 \right] }_{I_{\rm d}}
		- I_{\rm c}
		- \underbrace{ \frac{V + I R_{\rm s}}{R_{\rm sh}} }_{I_{\rm sh}}	
\end{equation}
where $I_{\rm ph}$ is the photocurrent;
$I_0$ is the diode reverse saturation current; 
$n$ is the diode ideality factor;
$V_{\rm th}=kT/q$ is the thermal voltage, with 
$q$ as the elementary charge ($1.602 \times 10^{-19}$C),
$k$ as the boltzmann constant ($1.38 \times 10^{-23}$J/K),
and $T$ as the cell temperature in Kelvins (K);
$R_{\rm s}$ and $R_{\rm sh}$ are the equivalent series and shunt resistances of the PV cell, respectively. 
$I_{\rm ph}$ and $I_0$ are functions of irradiance $G$ ($\text{W/m}^2$) and cell temperate $T$ (K), which can be calculated based on the performance model from the California Energy Commission (CEC) framework \cite{de2006improvement}. 
$I_{\rm c}$ is the charging current of the capacitor. Its diffusion capacitance $C$ is a function of its terminal voltage ($V_{\rm c} = V+I R_{\rm s}$), which is given by \cite{sharma1992determination}:
\begin{equation}\label{eq-cd}
\begin{split}
	C &= \frac{\tau}{nV_{\rm th}}I_{\rm 0} \exp \left( \frac{V_{\rm c}}{nV_{\rm th}} \right)
	= \frac{\tau}{nV_{\rm th}}(I_{\rm d} + I_{\rm 0})
\end{split}
\end{equation}
where $\tau$ is the minority-carrier lifetime (unit: second) that controls the strength of the hysteresis effect. Then, $I_{\rm c}$ can be expressed by:
\begin{equation}\label{eq-Ic}
\begin{split}
	I_{\rm c} &= \frac{{\rm d} (C\cdot V_{\rm c})  }{{\rm d} t} = C\frac{{\rm d}V_{\rm c}}{{\rm d} t} + 
	V_{\rm c}\frac{{\rm d}C}{{\rm d} t} \\
	&= C\frac{{\rm d}V_{\rm c}}{{\rm d} t} + V_{\rm c}\frac{{\rm d}C}{{\rm d} V_{\rm c}}
	\frac{{\rm d}V_{\rm c}}{{\rm d} t} \\
	&= \underbrace{\frac{\tau}{nV_{\rm th}}(I_{\rm d} + I_{\rm 0}) (1 + \frac{V_{\rm c}}{nV_{\rm th}})}_{C_{\rm eq}} 
	 \frac{{\rm d}V_{\rm c}}{{\rm d} t}
\end{split}
\end{equation}
where $C_{\rm eq}$ can be considered as the equivalent capacitance. 

\begin{figure}[htbp]
	\centering
	\includegraphics[width=1\columnwidth]{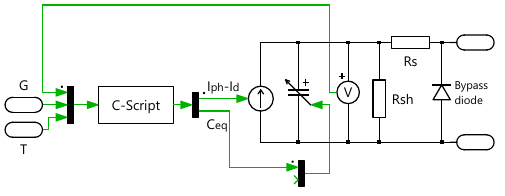}
	\caption{PV simulation model implemented by PLECS.}
	\label{fig-pv_model}
\end{figure}

Based on the cell model, we implement the PV simulation model in PLECS, as shown in Fig.~\ref{fig-pv_model}. The model uses a C-Script block to compute the $I_{\rm ph}-I_{\rm d}$ for the controllable current source, and $C_{\rm eq}$ for the controllable capacitor, based on the input irradiance $G$ and cell temperature $T$. To model the PV modules connected in series, we multiply the number of PV cell ($N_c$) on $R_{\rm s}$, $R_{\rm sh}$, and $V_{\rm th}$, respectively. Moreover, a bypass diode is also employed to replicate the real-world PV modules. 

\begin{figure}[htbp]
	\centering
	\includegraphics[width=1\columnwidth]{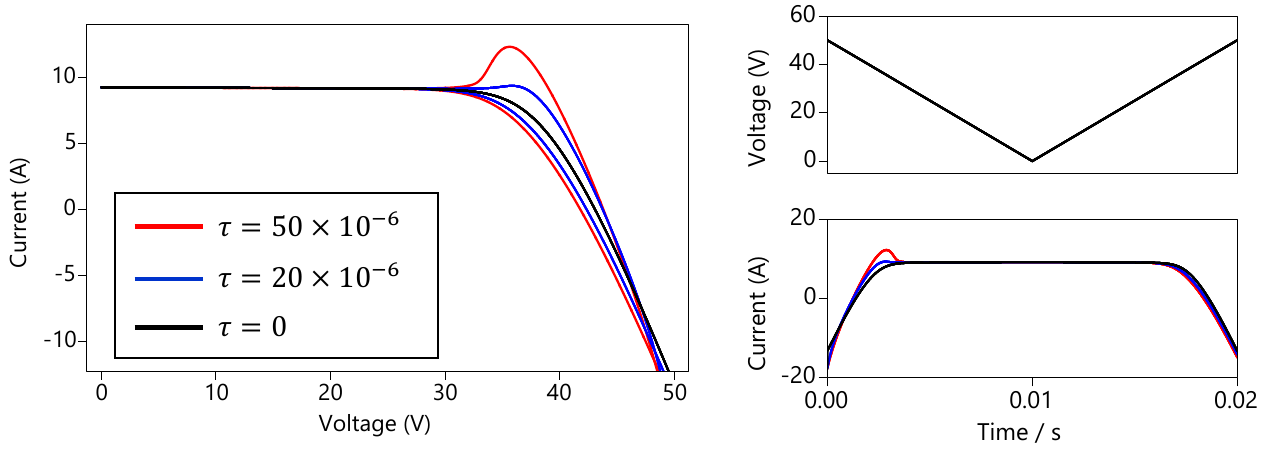}
	\caption{Dynamic characteristics of PV modules.}
		\vspace{-3mm}
	\label{fig-iv}
\end{figure}

Fig.~\ref{fig-iv} shows the simulated I-V characteristics of PV modules with different minority-carrier lifetime ($\tau = 0, 20\times10^{-6}, 50\times10^{-6}$). The terminal voltage of the PV modules is swept linearly from high to low (reverse scan) and from low to high (forward scan) within 0.02s. During the reverse scan, the actual current is higher than that of a static case ($\tau = 0$) since the internal capacitor is discharging. During the forward scan, the actual current will be lower than the static case since the internal capacitor is charging. Furthermore, PV modules with lager $\tau$ exhibit greater deviation, demonstrating a stronger hysteresis effect.

\begin{figure}[htbp]
	\centering
	\includegraphics[width=0.9\columnwidth]{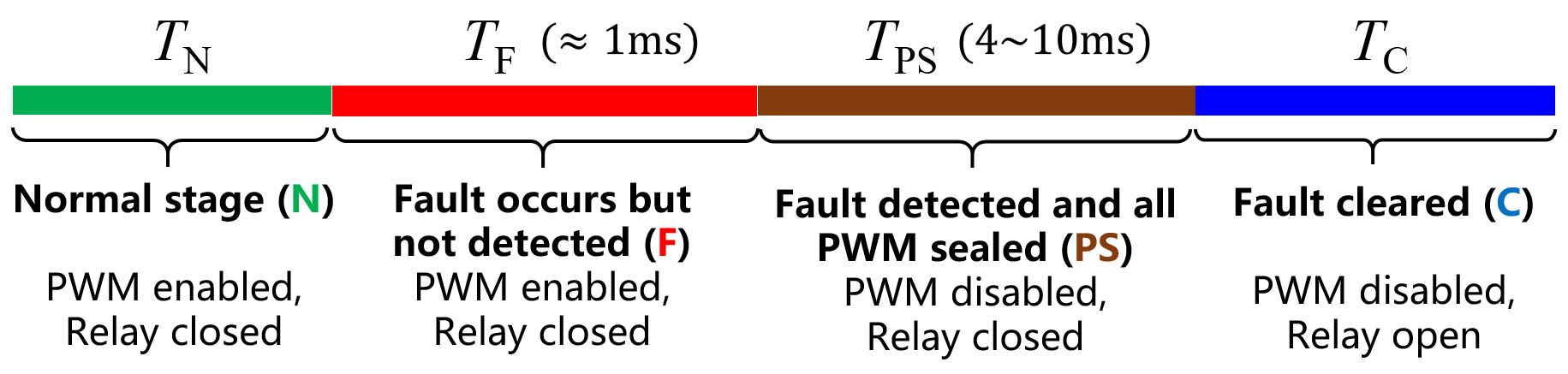}
	\caption{Four-stage shutdown procedure of PV inverter under GFs.}
	\label{fig-shutdown}
\end{figure}

\begin{figure*}[htbp] 
	\centering
	\subfloat[$N_{\rm sc}=0$\label{fig:exp_0sc}]{%
		\includegraphics[width=0.33\textwidth]{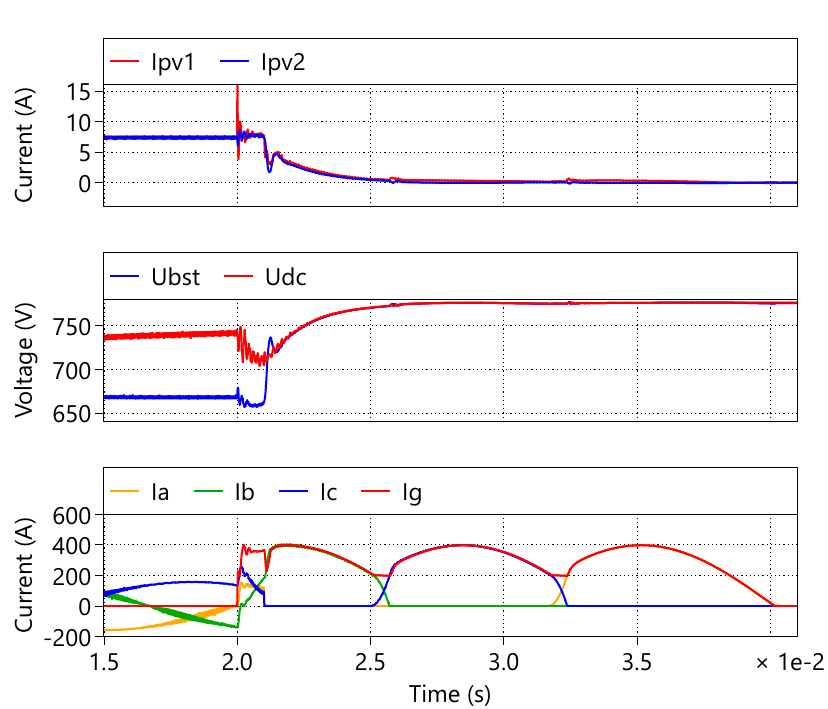}}
	\subfloat[$N_{\rm sc}=1$\label{fig:exp_1sc}]{%
		\includegraphics[width=0.33\textwidth]{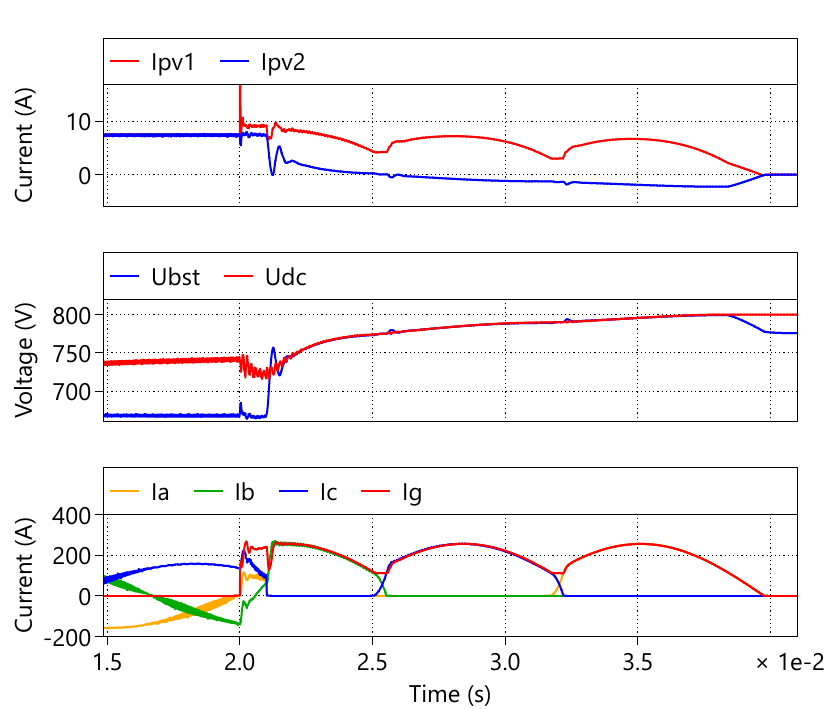}} 
	\subfloat[$N_{\rm sc}=6$\label{fig:exp_6sc}]{%
		\includegraphics[width=0.33\textwidth]{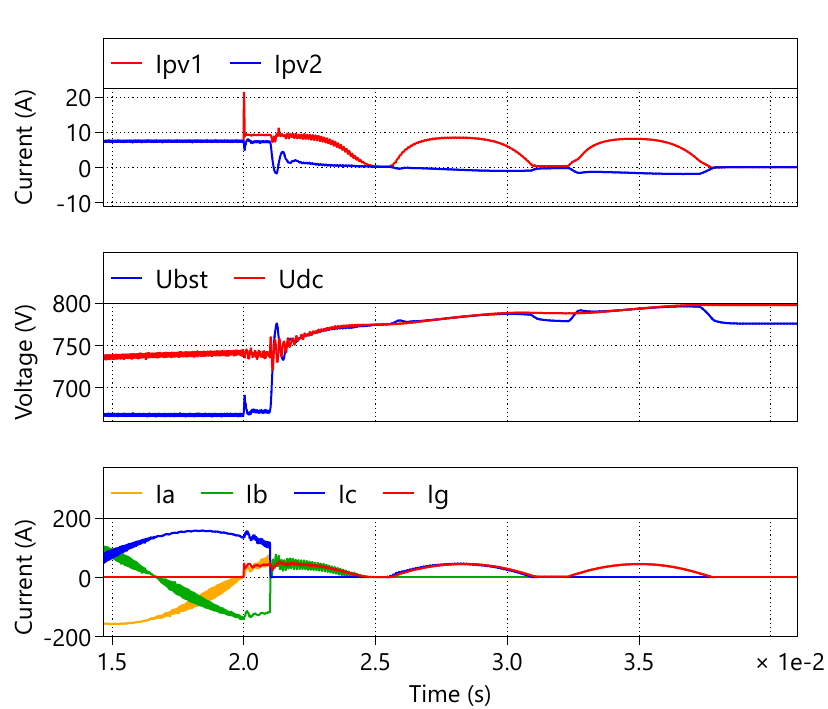}} \\[-14pt]
	\subfloat[$N_{\rm sc}=9$\label{fig:exp_9sc}]{%
		\includegraphics[width=0.33\textwidth]{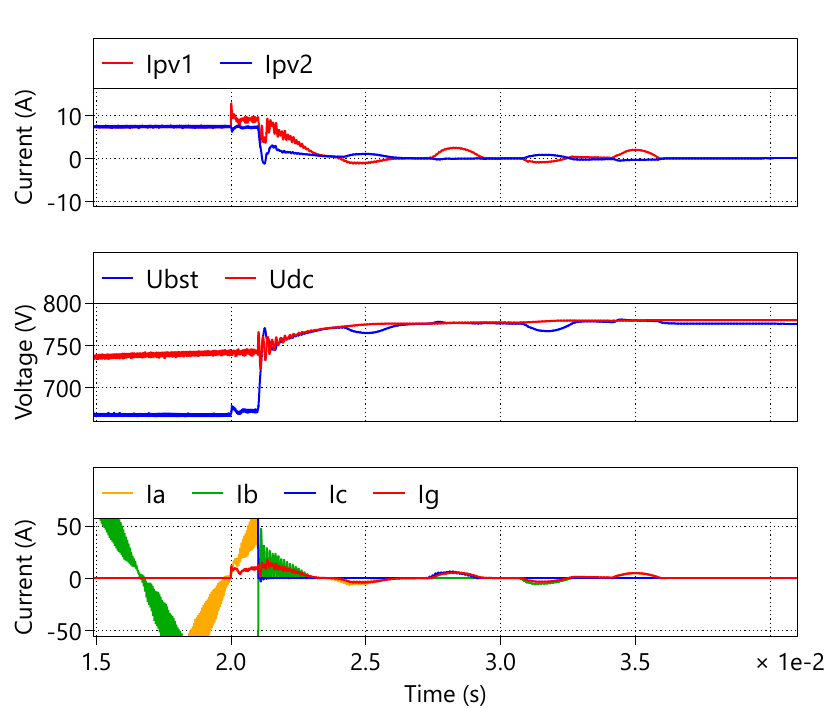}}
	\subfloat[$N_{\rm sc}=12$\label{fig:exp_12sc}]{%
		\includegraphics[width=0.33\textwidth]{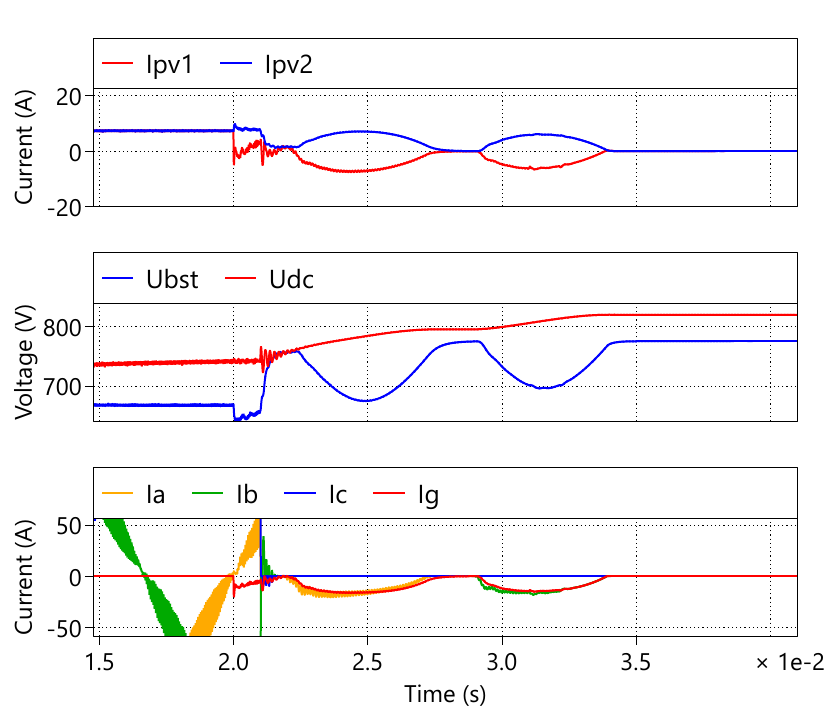}} 
	\subfloat[$N_{\rm sc}=18$\label{fig:exp_18sc}]{%
		\includegraphics[width=0.33\textwidth]{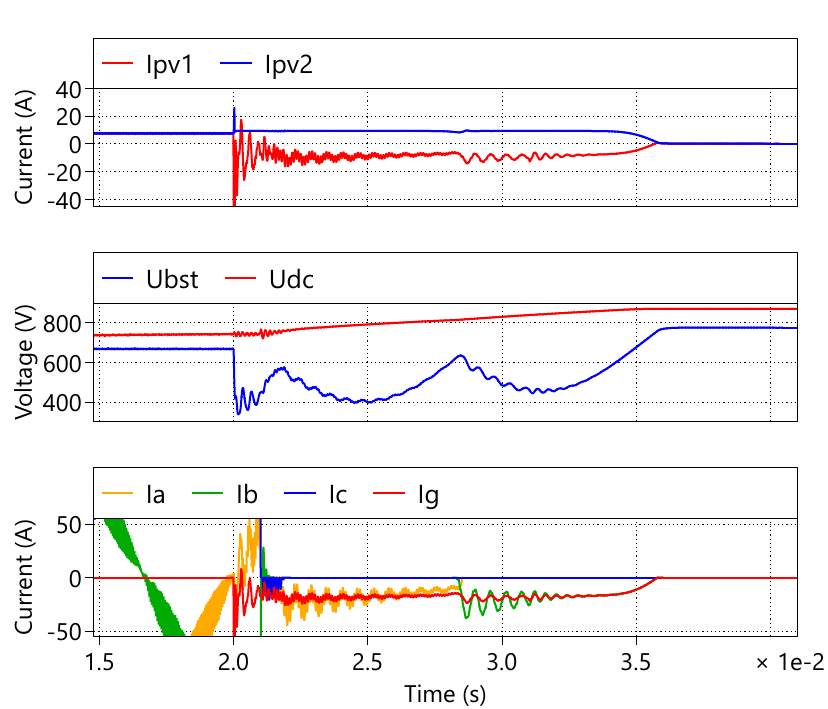}}
	\caption{Measurements of the PV inverter under GFs at different locations: $N_{\rm sc}=0, 1, 6, 9, 12, \text{and} 18$.}  
	\vspace{-2mm}
	\label{fig:ce-kl} 
\end{figure*}

\subsection{Four-Stage Shutdown Procedure} \label{subseciton-shutdown}
PV inverter’s shutdown procedure in the event of GF is modeled, which comprises four sequential stages, as illustrated in Fig.~\ref{fig-shutdown}. The first stage is the normal stage (N-stage), during which no fault has occurred. The system enters the second stage (F-stage) when a GF occurs, which will last for about 1 millisecond (ms) until the fault is detected by monitoring the zero-sequence current, i.e., $I_{\rm g}$. After that, all PWM signals are disabled in the third stage (PS-stage), and a trip command is issued to the relay at the same time. After a delay of roughly 4-10 ms, the circuit break opens to clear the fault, which enters the final stage (C-stage).

The measurements captured during the shutdown procedure are used for post-analysis. We assume the time duration of the recorded shutdown procedure is $T$, with the duration of each stage denoted as $T_{\rm N}$, $T_{\rm F}$, $T_{\rm PS}$, and $T_{\rm C}$.

\section{Case Study} \label{section-case_study}
This section presents typical GF cases based on the proposed simulation model, where fault locations are varied from the negative terminal to the positive terminal by adjusting the number of short-circuited PV modules ($N_{\rm sc}$). Specifically, we model a 1 kV-level PV system using $N_{\rm pv}=18$ PV modules per string, and we design six cases with $N_{\rm sc}=$ 0, 1, 6, 9, 12, and 18. For each case, we set $T_{\rm N}=20$ ms, $T_{\rm F}=1$ ms, $T_{\rm PS}=10$ ms, and $T_{\rm C}=10$ ms. GFs occur in the $1^{\text{st}}$ string of the $1^{\text{st}}$ boost converter under the irradiance of 1000 $\text{W/m}^2$.

\subsection{At Negative Terminal $(N_{\rm sc}=0)$}
Fig.~\ref{fig:exp_0sc} shows the measurements from the PV inverter under $N_{\rm sc}=0$. Since the GF is very close to the common negative terminal within the inverter, the post-fault distance between two string currents ($I_{\rm pv1}$ and $I_{\rm pv2}$) is very small with $I_{\rm pv1}$ being slighter higher than $I_{\rm pv2}$. Some high-frequency components can be observed from  $I_{\rm pv1}$ in the F-stage. During the PS-stage, large lower-leg fault currents flow through each phase ($I_{\rm a}$, $I_{\rm b}$, $I_{\rm c}$), resulting in a large positive fault current ($I_{\rm g}$). Both $U_{\rm bst}$ and $U_{\rm dc}$ will converge to PV string's open-circuit voltage ($U_{\rm oc}$) after the fault is cleared. 

\subsection{Within String and Close to Negative Terminal $(N_{\rm sc}=1, 6)$}
\begin{figure}[htbp] 
	\centering
	\includegraphics[width=1\columnwidth]{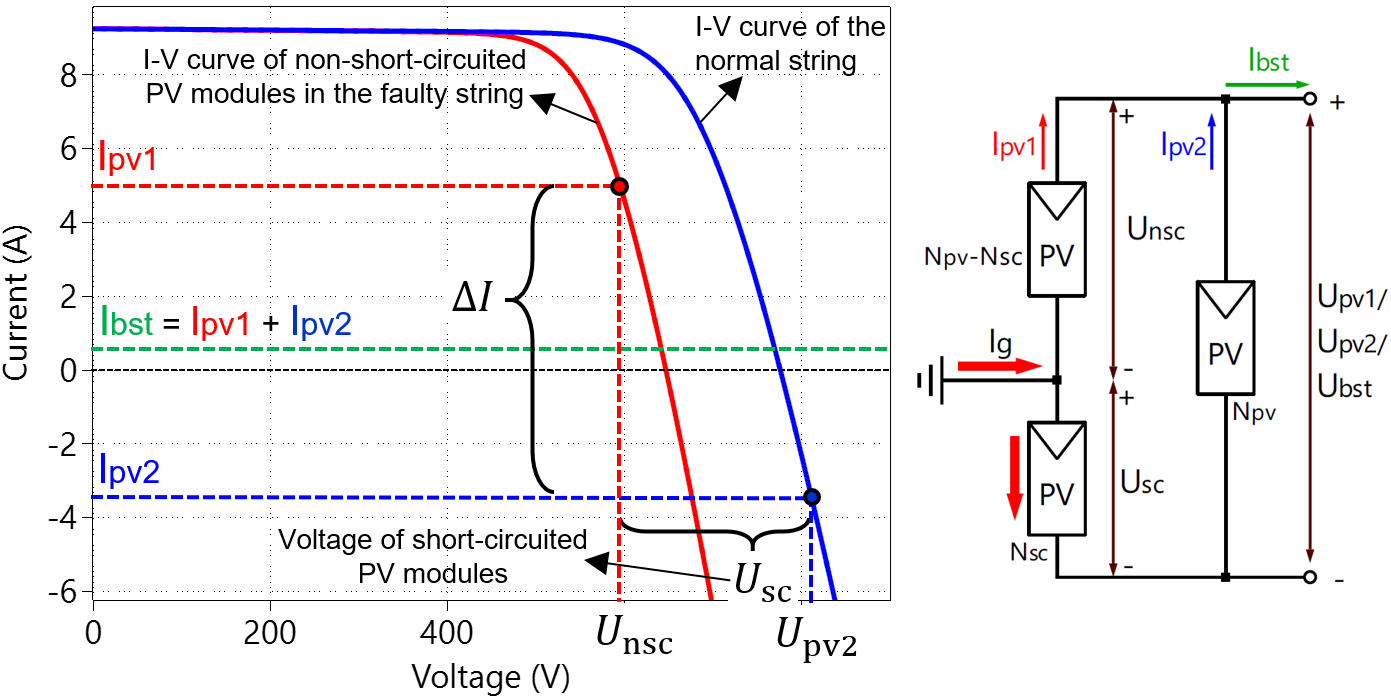}
	\caption{The cause of current difference between strings under GFs.}
		\vspace{-3mm}
	\label{fig-faulty_iv}
\end{figure}

When a small number of PV modules are short-circuited by a GF, an obvious post-fault current difference $\Delta I = I_{\rm pv1}-I_{\rm pv2}$ can be observed, as shown in Figs.~\ref{fig:exp_1sc} and \ref{fig:exp_6sc}. This can be generally explained by Fig.~\ref{fig-faulty_iv}, where the red I-V curve denotes the trajectory of the non-short-circuited PV modules in string-1, and the blue curve represents string-2. The short-circuited modules act as a resistance that creates a voltage drop $U_{\rm sc}$ in string-1, which leads to $\Delta I$ by the constrains $I_{\rm bst} = I_{\rm pv1} + I_{\rm pv2}$ and $U_{\rm pv2} = U_{\rm nsc} + U_{\rm sc}$. Since $U_{\rm sc}$ is produced by $I_{\rm g}$, $I_{\rm pv1}$ is correlated with $I_{\rm g}$.

$\Delta I$ in the PS-stage is more pronounced since PV's operating point moves toward the open-circuit region, where the slope of the I–V curve is steeper compared to the MPPT region in the F-stage; thus, small voltage drops cause comparatively larger current shifts. 

Fig.~\ref{fig:exp_6sc} presents a lower $I_{\rm g}$ compared to Fig.~\ref{fig:exp_1sc} since more PV modules are short-circuited, which brings a higher impedance into the fault-current loop.

$U_{\rm bst}$ and $U_{\rm dc}$ exceed $U_{\rm oc}$ in the PS-stage since string-2 operates in the 4$^{\text{th}}$ quadrant of the I-V curve, i.e., $I_{\rm pv2}<0$. When the fault is cleared, $U_{\rm bst}$ decreases back to $U_{\rm oc}$ while $U_{\rm dc}$ retains the elevated voltage due to the absence of a discharging loop for DC-link capacitors.

\subsection{In the Middle $(N_{\rm sc}=9)$}
As shown in Fig.~\ref{fig:exp_9sc}, when half of PV modules are short-circuited by a GF, lower-leg fault currents continue to decrease while upper-leg fault currents begin to emerge. As a result, $I_{\rm g}$ oscillates around zero with a small amplitude, while $I_{\rm pv1}$ and $I_{\rm pv2}$ change symmetrically. Under this mid-string location, the negative impact of the GF on the PV inverter is relatively lower other locations.

\subsection{Within String and Close to Positive Terminal $(N_{\rm sc}=12)$}
As the fault moves further toward the positive terminal, upper-leg fault currents increase and lower-leg fault currents vanish, which results in a negative fault current and an increase in $U_{\rm dc}$, as shown in Fig.~\ref{fig:exp_12sc}. Meanwhile, $I_{\rm pv1}$ will be negative and lower than $I_{\rm pv2}$. 

\subsection{At Positive Terminal $(N_{\rm sc}=18)$}
When all PV modules are short-circuited, the upper-leg fault currents reach their maximum magnitude, producing the largest negative fault current and the highest $U_{\rm dc}$ (see Fig.~\ref{fig:exp_18sc}). This is because the GF at this location causes all PV modules in string-1 to work together with the grid source to charge the DC-link capacitors.

From the above case studies, the characteristics of GF vary significantly with fault locations, and both the faulty and normal string currents deviate during the F- and PS-stages, making the localization difficult by using just one or few features. However, some features remain informative. For example, the normal string current correlates better with $U_{\rm bst}$, while the faulty string current shows better correlation with $U_{\rm dc}$ and $I_{\rm g}$. Accordingly, correlation-based features are adopted in the GF localization method, which will be introduced in Section~\ref{subsection-feature-extraction}.

\section{Ground Fault Localization Model} \label{section-ai}
\begin{figure}[htbp]
	\centering
	\includegraphics[width=1\columnwidth]{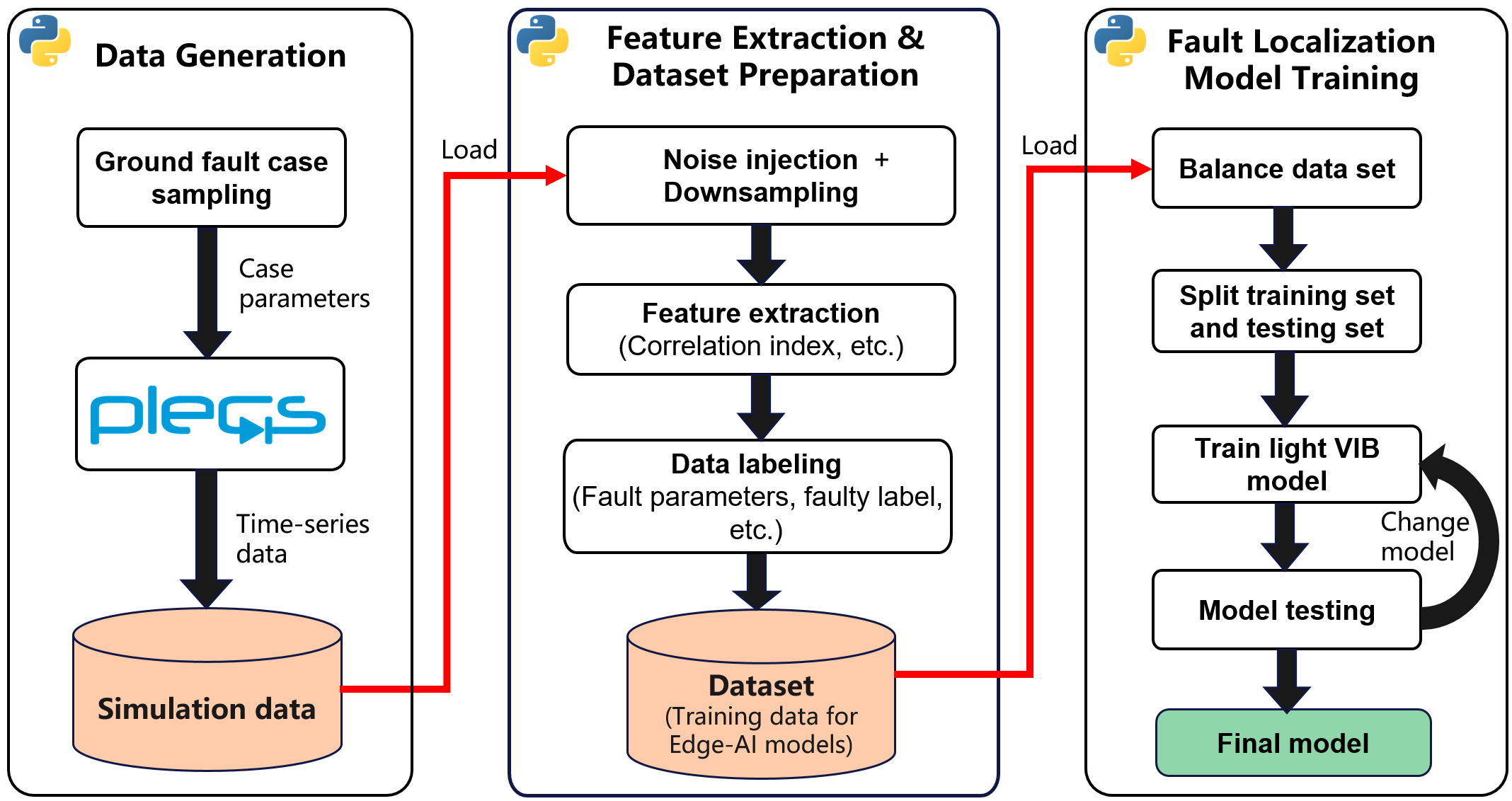}
	\caption{Edge-AI development pipeline for GF localization.}
		\vspace{-3mm}
	\label{fig-ai_framework}
\end{figure}

Based on the proposed simulation model built in PLECS, we can generate a comprehensive dataset of diverse GF cases to train a localization model for edge-side PV inverters. Specifically, we develop an offline edge-AI development pipeline, as shown in Fig. \ref{fig-ai_framework}, which consists of three phases: (1) data generation, (2) feature extraction and dataset preparation, and (3) fault localization model training. The localization model is based on the Variational Information Bottleneck (VIB) model to determine whether a PV string has a GF at the negative terminal, or at other locations, or no fault at all.

\subsection{Data Generation}\label{subsection-data-generation}
In the data generation phase of the pipeline, a batch of GF simulation cases will be generated under diverse GF configurations. Each case is configured by a set of parameters, including irradiance, cell temperature, minority-carrier lifetime of PV modules, GF location (adjusted by the number of short-circuited PV modules), ground resistance, F-stage duration, and PS-stage duration. All parameters are sampled uniformly from predefined ranges, as summarized in Table~\ref{tab:gf_parameters}. For each simulation case, we record all measurable signals (as highlighted in Fig. \ref{fig-inverter_circuit}) under 100 kHz sampling frequency. 
\begin{table}[htbp]
	\centering
	\caption{Key Parameters of GF Cases}
	\label{tab:gf_parameters}
	\begin{tabular}{cccc}
		\toprule
		\textbf{Name} & \textbf{Symbol} & \textbf{Range} / \textbf{Value}  & \textbf{Unit, Type} \\
		\midrule
		Irradiance & $G$ & [100, 1200] & $\text{W/m}^2$, Float \\
		Cell temperature & $T$ & [320, 350] & K, Float \\
		Minority-carrier lifetime & $\tau$ & [10, 50] & us, Float\\
		GF location & $N_{\rm sc}$ & [0, 18] & Integer \\
		Ground resistance & $R_{\rm g}$ & [0.8, 2.0] & $\Omega$, Float \\
		F-stage duration & $T_{\rm F}$ & [0.6, 1.0] & ms, Float\\
		PS-stage duration & $T_{\rm PS}$ & [4, 10] & ms, Float\\
		\bottomrule
	\end{tabular}
\end{table}

\subsection{Feature Extraction and Dataset Preparation} \label{subsection-feature-extraction}

\begin{figure*}[bt] 
	\centering
	\subfloat[$I_{\rm pv1}$\label{fig:decom_Ipv1}]{%
		\includegraphics[width=0.33\textwidth]{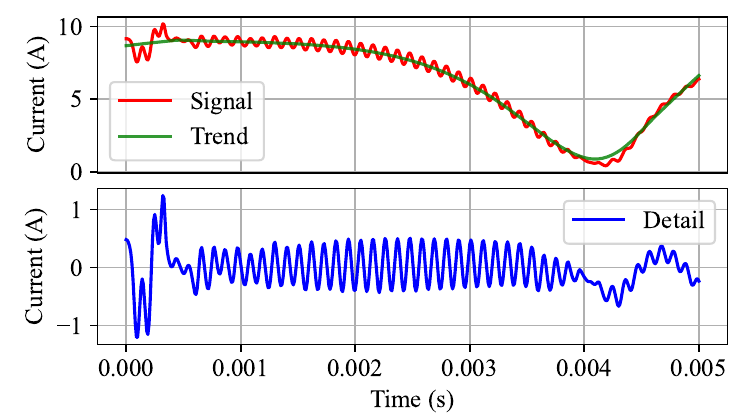}}
	\subfloat[$U_{\rm iso}$\label{fig:decom_Uiso}]{%
	\includegraphics[width=0.33\textwidth]{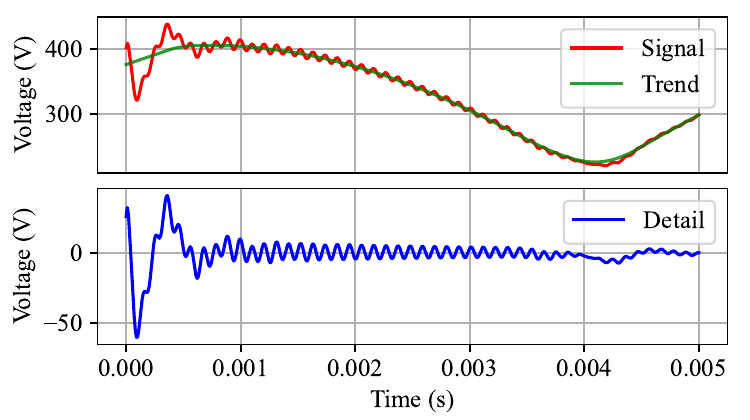}} 
	\subfloat[$I_{\rm g}$\label{fig:decom_I0}]{%
	\includegraphics[width=0.33\textwidth]{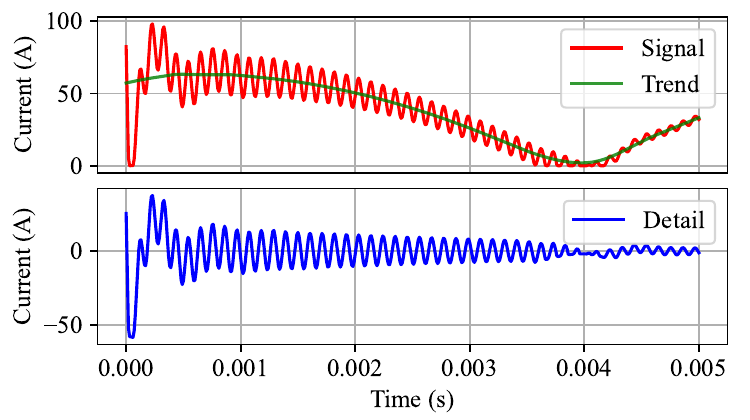}}  \\[-8pt]
	\subfloat[$I_{\rm pv2}$\label{fig:decom_Ipv2}]{%
	\includegraphics[width=0.33\textwidth]{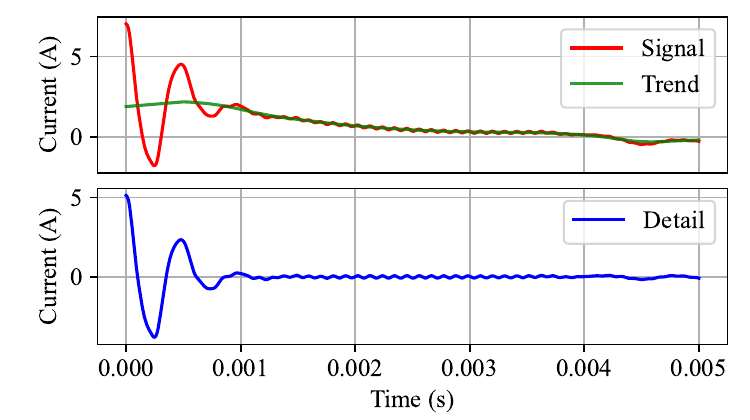}}
	\subfloat[$U_{\rm bst}$\label{fig:decom_Ubst}]{%
		\includegraphics[width=0.33\textwidth]{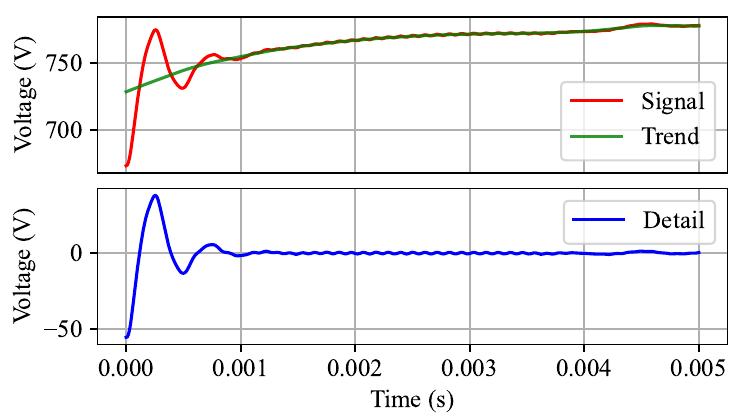}} 
	\subfloat[$U_{\rm dc}$\label{fig:decom_Udc}]{%
		\includegraphics[width=0.33\textwidth]{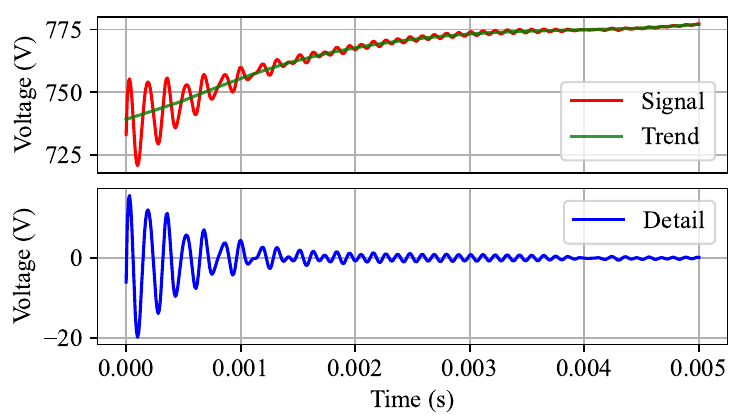}} 
	\caption{LOESS-based signal decomposition on $I_{\rm pv1}$, $U_{\rm iso}$, $I_{\rm g}$, $I_{\rm pv2}$, $U_{\rm bst}$, and $U_{\rm dc}$, under $N_{\rm sc}=5$.}  
	\label{fig:decom} 
    \vspace{-2mm}
\end{figure*}

The second phase of the pipeline will prepare a structured dataset by processing the time-series simulation data. First, it injects noise with multiple scales into the time-series and performs down-sampling to emulate practical scenarios. Next, it performs feature extraction on three PV strings in each simulation case: (1) the faulty string, (2) a normal string under the faulty boost, and (3) a normal string under a normal boost. Each string is assigned with a label: `0' for the normal string, `1' if it has a GF at the negative terminal ($N_{sc}=0$), and `2' if it has a GF with $N_{sc}>0$. We treat category `1' separately because faults at the negative terminal are relatively difficult to recognize and have a more severe impact on the inverter than faults at other locations.

Since the GF characteristics differ between the F-stage and PS-stage, feature extraction is performed separately in these two stages for each PV string. 
The features primarily consist of correlation measures between the string current ($I_{\rm pv}$) and other shared signals ($U_{\rm bst}$, $U_{\rm dc}$, $U_{\rm iso}$, and the estimated fault current $I_{\rm g} \approx I_{\rm a} + I_{\rm b} + I_{\rm c}$). This is because correlation metrics are unitless, insensitive to variations in magnitude, and normalized to range -1 to +1, which can improve the generalizability of the localization model.

The faulty string current is affected by both the fault-current loop and the string’s inherent I-V characteristics. Its high-frequency component primarily arises from the fault current that causes dynamic changes in $U_{\rm sc}$, as illustrated in Fig.~\ref{fig-faulty_iv}. Therefore, to improve the correlation-based feature extraction, signal decomposition based on the LOESS (Locally Estimated Scatterplot Smoothing) \cite{cleveland1979robust} algorithm is applied to extract the trend component and the detail component (high-frequency component obtained by subtracting the trend from the original signal) of each signal. An example under the GF with $N_{\rm sc}=5$ is illustrated in Fig.~\ref{fig:decom}. In this case, the detail component of the faulty string current ($I_{\rm pv1}$) exhibits strong correlations with the detail component of $I_{\rm g}$ and $U_{\rm iso}$. Meanwhile, the detail component of the normal string current ($I_{\rm pv2}$) has stronger negative correlation with the detail component of $U_{\rm bst}$. In addition, different GF locations will result in different correlation pattern between trend components. Fig.~\ref{fig:decom} presents one case in which the trend of $I_{\rm pv1}$, $I_{\rm g}$, and $U_{\rm iso}$ are well-correlated. However, when $N_{\rm sc}=0$, such trend correlation weakens due to the week $U_{\rm sc}$ and the dominant influence becomes the string's inherent I-V characteristics. Under such condition, the trend components of both $I_{\rm pv1}$ and $I_{\rm pv2}$ will show a strong negative correlation with $U_{\rm bst}$, but remain distinguishable by analyzing their detail components.

Considering the non-linear behavior of PV systems, we employ two types of correlation coefficients: the Pearson correlation ($r_{\rm p}$) to capture linear relationships \cite{pearson_wiki} and the Spearman correlation ($r_{\rm s}$) to capture nonlinear relationships \cite{spearman_wiki}, as expressed in \eqref{eq:pearson} and \eqref{eq:spearman}, respectively.

\begin{equation}\label{eq:pearson}
	r_{\rm p} = \frac{\displaystyle \sum_{i=1}^n (x_i - \bar x)\,(y_i - \bar y)}%
	{\sqrt{\displaystyle \sum_{i=1}^n (x_i - \bar x)^2}\;\sqrt{\displaystyle \sum_{i=1}^n (y_i - \bar y)^2}}
\end{equation}
\begin{equation}\label{eq:spearman}
	r_{\rm s} = 1 - \frac{6 \sum_{i=1}^n d_i^2}{n\,(n^2 - 1)}
\end{equation}
where $x_i$ and $y_i$ are the $i^\text{th}$ observations of the two signals, $n$ is the number of data points, and $d_i$ is the rank difference between $x_i$ and $y_i$. Table~\ref{tab:features} lists all correlation-based features from F-stage or PS-stage (24 per stage), where $r(\cdot)$ denotes a kind of correlation coefficient. The subscripts $\rm p$ and $\rm s$ denote the Pearson and Spearman correlations, respectively, while the superscripts $\rm o$, $\rm d$, and $\rm t$ indicate correlations computed from the original signals, the detail components, and the trend components, respectively.

\begin{table}[htbp]
	\centering
	\caption{Correlation-based features from F-stage or PS-stage}
	\label{tab:features}
	\begin{tabular}{cc}
		\toprule
		\textbf{Signal pair} & \textbf{Correlation index} \\
		\midrule
		$I_{\rm pv}$ vs. $U_{\rm bst}$ & 
		$r_{\{\rm p,s\}}^{\{\rm o,d,t\}}(I_{\rm pv},\,U_{\rm bst})$ \\
		$I_{\rm pv}$ vs. $U_{\rm dc}$ & 
		$r_{\{\rm p,s\}}^{\{\rm o,d,t\}}(I_{\rm pv},\,U_{\rm dc})$   \\
		$I_{\rm pv}$ vs. $U_{\rm iso}$ & 
		$r_{\{\rm p,s\}}^{\{\rm o,d,t\}}(I_{\rm pv},\,U_{\rm iso})$  \\
		$I_{\rm pv}$ vs. $I_{\rm g}$ & 
		$r_{\{\rm p,s\}}^{\{\rm o,d,t\}}(I_{\rm pv},\,I_{\rm g})$ \\
		\bottomrule
	\end{tabular}
\end{table}

In addition, the mean value of the string current is also included in the feature set. Finally, a total of 50 features are extracted from the two stages for each PV string. Let $M$ denote the number of simulation cases. The dataset prepared at the end of this phase will contain $3M$ samples (records), with each sample comprising 50 features and one label.

\subsection{Fault Localization Model} \label{subsection-fault-localization}
In the third phase of the pipeline, a fault localization model based on the Variational Information Bottleneck (VIB) is developed and trained. VIB originates from the Information Bottleneck (IB) theory \cite{tishby2000information}, which seeks an encoding $Z$ of the input $X$ that retains maximal information about the target $Y$ while discarding irrelevant information from $X$ (maximal compression). Formally, the IB objective is given by: 

\begin{equation}\label{eq-ib}
	\mathcal{J}_{\rm IB} = I(Z;Y) - \beta I(X;Z)
\end{equation}
where $I(\cdot)$ denotes the mutual information between two random variables, and $\beta$ trades off between the representational and compression capability of $Z$. VIB implements the IB objective by using a encoder-decoder framework based on artificial neural networks (ANNs) \cite{Alemi2017DeepVI}. The encoder acts as a Gaussian encoder that outputs distribution parameters of the latent variable $\mathbf{z}$ (encoding). The decoder maps $\mathbf{z}$ to a target through an ANN. During the training phase, $\mathbf{z}$ is sampled from the Gaussian distribution. During inference,  $\mathbf{z}$ is set to the mean of the trained Gaussian distribution. 

\begin{figure}[htbp]
	\centering
	\includegraphics[width=0.9\columnwidth]{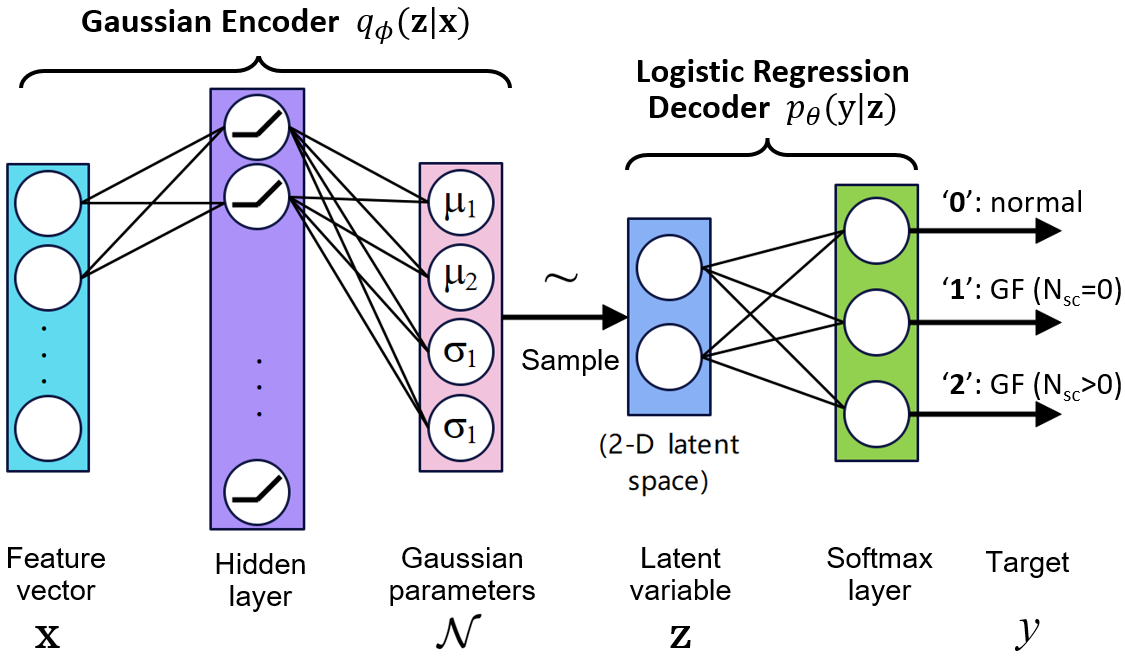}
	\caption{The lightweight VIB model for GF localization.}
	\label{fig-vib}
\end{figure}

Based on this framework, we design a lightweight VIB architecture as the fault-localization model (see Fig.~\ref{fig-vib}), enabling the deployment on edge-side PV inverters. Its input is a feature vector $\mathbf{x}$ of a PV string. The Encoder $q_\phi(\mathbf{z}|\mathbf{x})$, parameterized by $\phi$, is a 2-d Gaussian encoder represented by an ANN with one hidden layer. Thus, $\mathbf{z}$ is sampled from a 2-d latent-space. The decoder $p_\theta(y |\mathbf{z})$, parameterized by $\theta$, is a multi‐output logistic-regression model comprising a fully connected layer followed by a Softmax layer, which outputs the probability of each label ($y$): `0' (normal string), `1' (GF with $N_{\rm sc}=0$), and `2': (GF with $N_{\rm sc}>0$). 

Assume a training set $\mathcal{D}$ contains $N$ labeled samples, each denoted as $(\mathbf{x}, y) \in \mathcal{D}$. The loss function to be minimized is formulated as \cite{Alemi2017DeepVI}:
\begin{equation}
\begin{split}
	\mathcal{L}(\theta,\phi;\mathcal{D}) = &\frac{1}{N}\sum_{(\mathbf{x}, y)\in\mathcal{D}}\Bigl\{  - \mathbb{E}_{\varepsilon\sim\mathcal N}\left[\log p_\theta(y \mid f(\mathbf{x},\epsilon))\right] \\
	&+ \beta \, D_\mathrm{KL}\bigl(q_\phi(Z \mid \mathbf{x})\,\|\,\mathcal{N}(\mathbf{0}, \mathbf{I})\bigr) \Bigr\}
\end{split}
\end{equation}
where the first term maximizes the expected log‐likelihood (i.e., representational performance). Here, $f(\mathbf{x},\epsilon)$ denotes a differentiable sampling process of $\mathbf{z}$ based on the reparameterization tick \cite{kingma2019introduction}, i.e., $\mathbf z = \mu_{\phi}(\mathbf x) + \sigma_{\phi}(\mathbf x)\cdot\varepsilon$, with $\varepsilon\sim\mathcal N(0,1)$. 
The second term tries to reduce the Kullback–Leibler (KL) divergence between the latent‐space distribution produced by the encoder and the standard normal prior, thereby regulating the encoder’s compression capability and preventing overfitting.

Finally, by applying a gradient-decent-based optimizer, such as AdamW \cite{Loshchilov2017FixingWD}, to minimize this loss function over the dataset, the string‐level GF localization model can be obtained.

\section{Experimental Validation} \label{section-experiment}
From the data generation phase, over 500 simulation cases covering various fault configurations were obtained. Through noise injection with multiple scales, downsampling, and data balancing, we finally constructed three datasets at sampling rates of 100 kHz, 10 kHz, and 5 kHz, each containing 7160 samples. In each dataset, 50\% of the samples are labeled as the normal strings (`0'), half of which stem from the faulty boost and the other half from the normal boost; 25\% are labeled as GF with $N_{\rm sc}=0$ (`1'); and the remaining 25\% are labeled as GF with $N_{\rm sc}>0$ (`2'). For each dataset, 80\% of the samples are used for training, while the remaining are used for testing. 

The proposed lightweight VIB model is implemented by Pytorch. Its training configurations and hyperparameters are listed in Table~\ref{tab:yyperparameters}, which are empirically determined via multiple experiments.

\begin{table}[htbp]
	\centering
	\caption{Hyperparameters of the Model Training Process}
	\label{tab:yyperparameters}
	\begin{tabular}{ccc}
		\toprule
		\textbf{Parameter} & \textbf{Value}   \\
		\midrule
		Number of neuros in the hidden layer & 12 \\
		Latent-space dimension & 2\\
		KL penalty coefficient ($\beta$) & 0.01 \\
		Learning rate & 0.001 \\
		Batch size & 1000 \\
		Number of training epoches & 4000\\
		\bottomrule
	\end{tabular}
\end{table}

\begin{figure*}[htbp] 
	\centering
	\subfloat[CE loss (all features)\label{fig:ce}]{%
		\includegraphics[width=0.5\columnwidth]{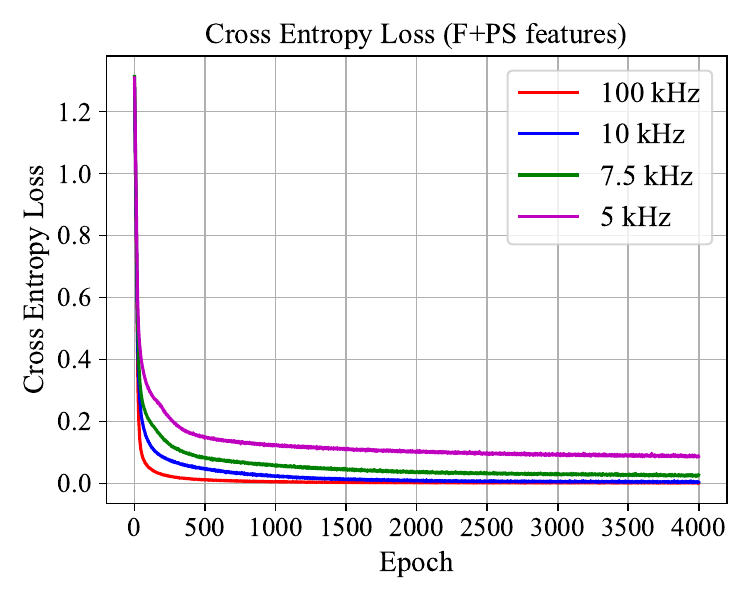}}
	\subfloat[CE loss (PS features)\label{fig:ce_is}]{%
		\includegraphics[width=0.5\columnwidth]{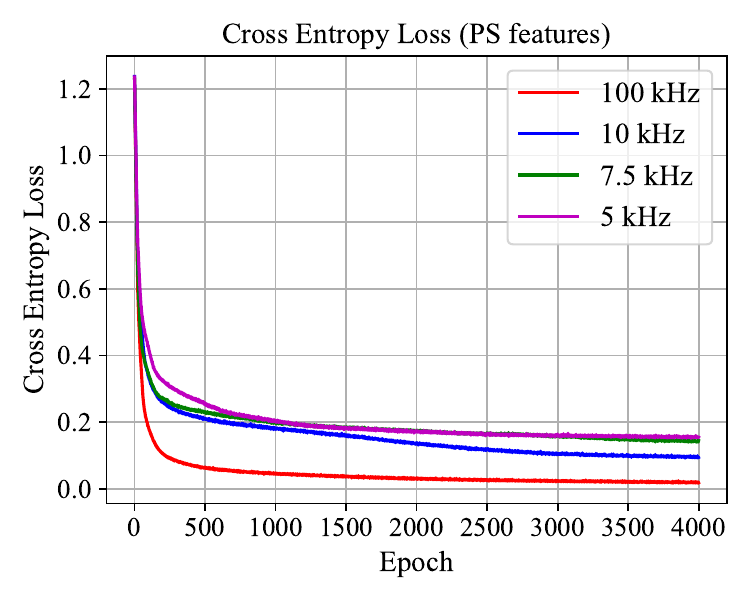}}
	\subfloat[KL divergence (all features)\label{fig:kl}]{%
		\includegraphics[width=0.5\columnwidth]{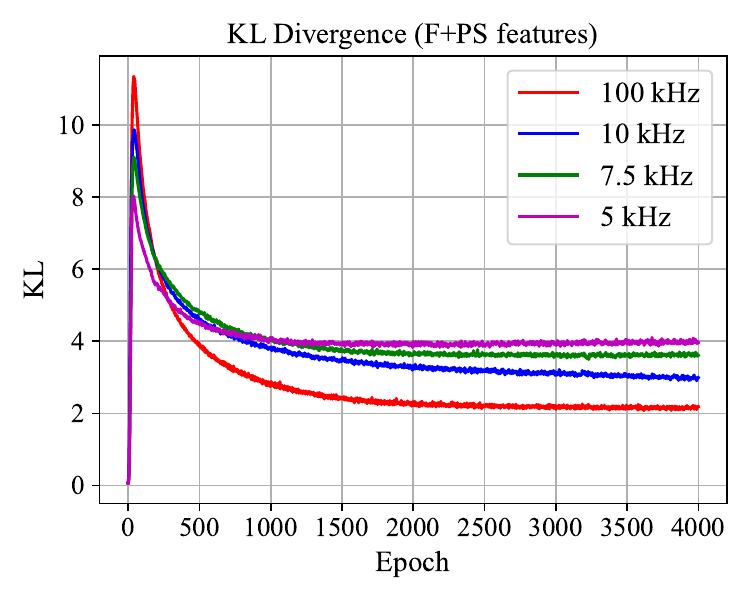}}
	\subfloat[KL loss (PS features)\label{fig:kl_is}]{%
		\includegraphics[width=0.5\columnwidth]{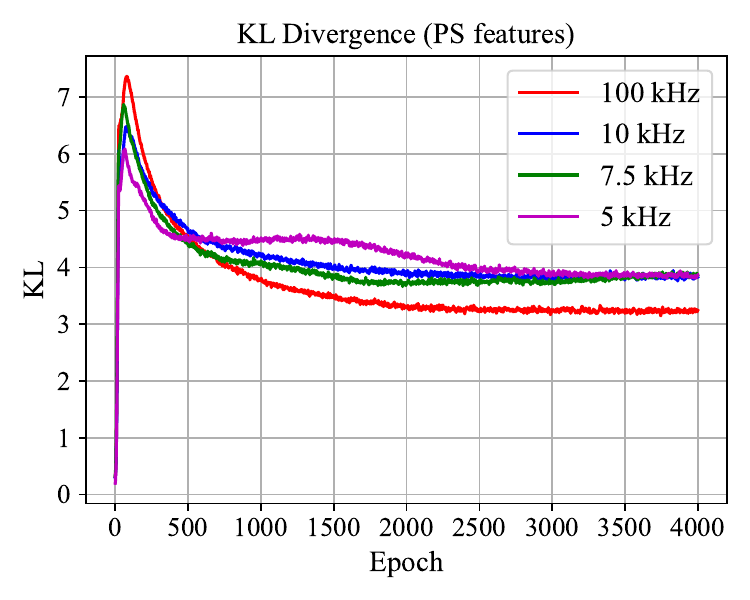}}
	\caption{CE loss and KL divergence of training under four sampling rates and two feature sets.}  
	\label{fig:ce-kl} 
\end{figure*}

\begin{figure*}[htbp] 
	\centering
	\subfloat[100kHz, all features\label{fig:latent-space-100}]{%
		\includegraphics[width=0.5\columnwidth]{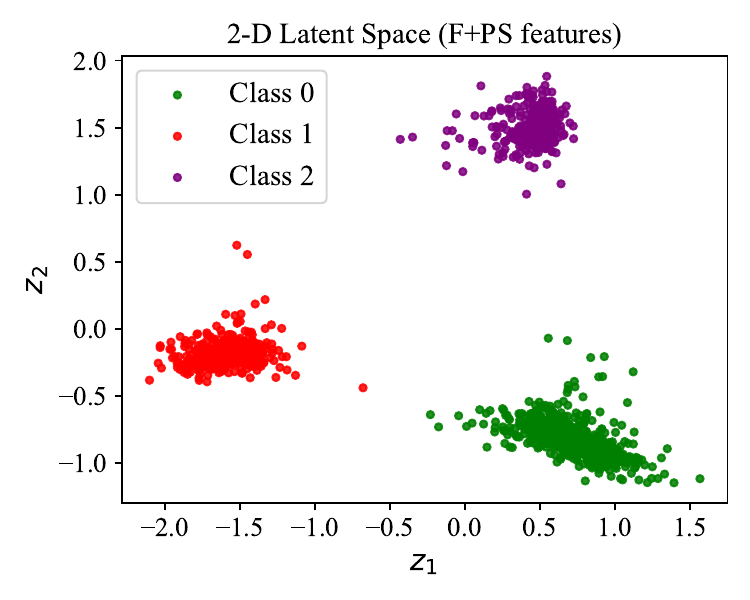}}
	\subfloat[10kHz, all features\label{fig:latent-space-10}]{%
		\includegraphics[width=0.5\columnwidth]{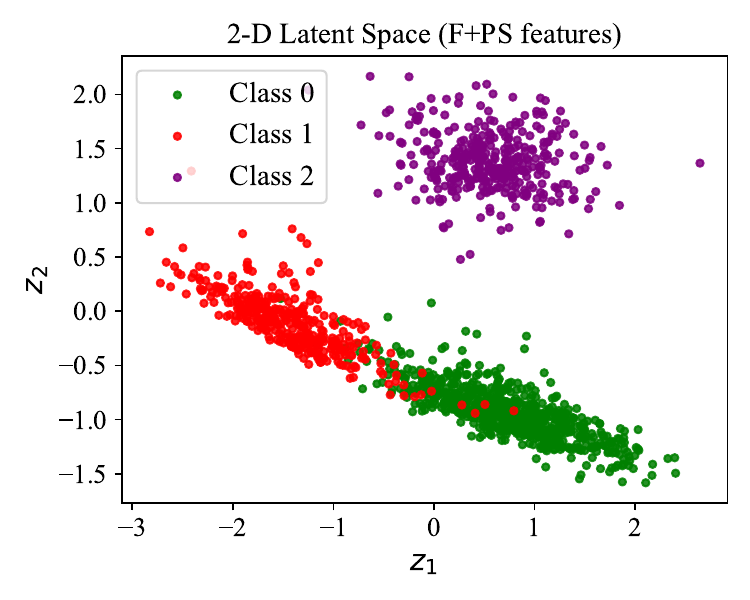}}
	\subfloat[7.5kHz, all features\label{fig:latent-space-7-5}]{%
		\includegraphics[width=0.5\columnwidth]{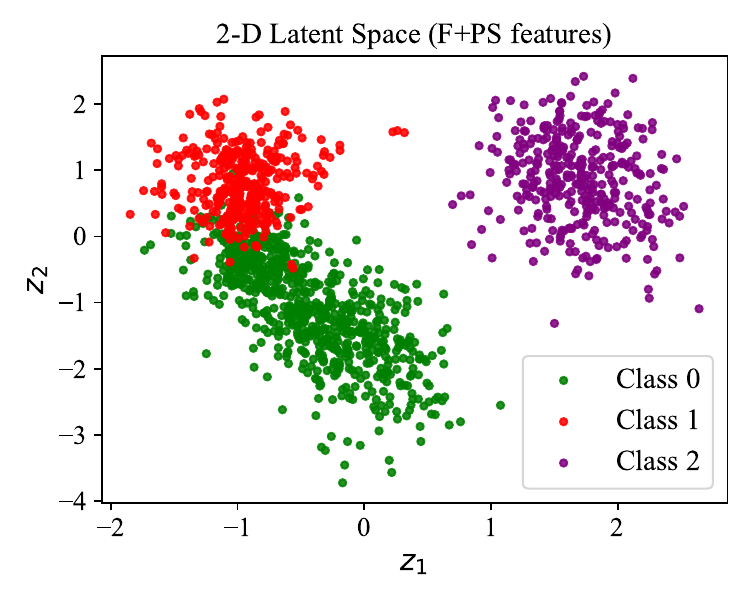}}
	\subfloat[5kHz, all features\label{fig:latent-space-5}]{%
		\includegraphics[width=0.5\columnwidth]{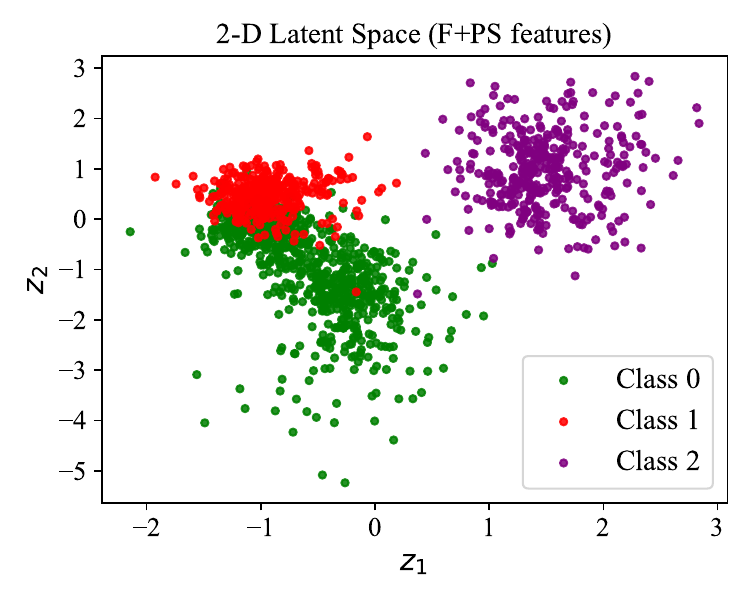}} \\ [-8pt]
	\subfloat[100kHz, PS features\label{fig:latent-space-100_is}]{%
		\includegraphics[width=0.5\columnwidth]{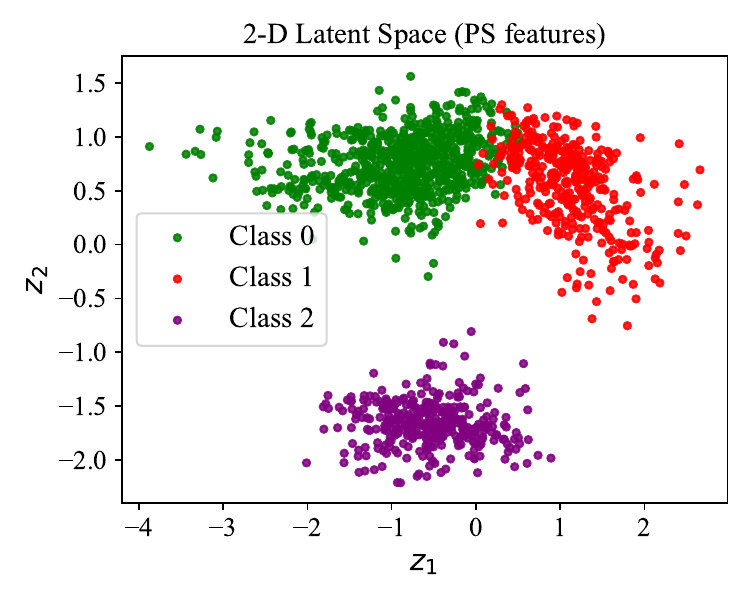}}
	\subfloat[10kHz, PS features\label{fig:latent-space-10_is}]{%
		\includegraphics[width=0.5\columnwidth]{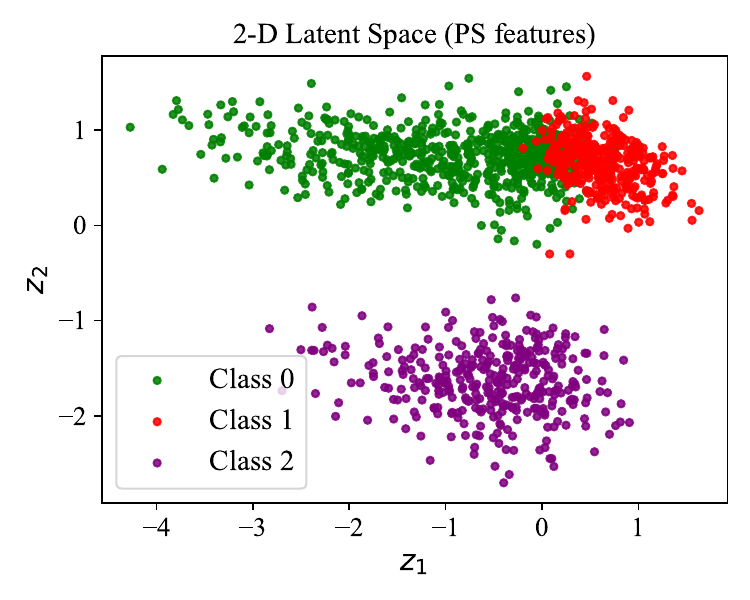}}
	\subfloat[7.5kHz, PS features\label{fig:latent-space-7-5_is}]{%
		\includegraphics[width=0.5\columnwidth]{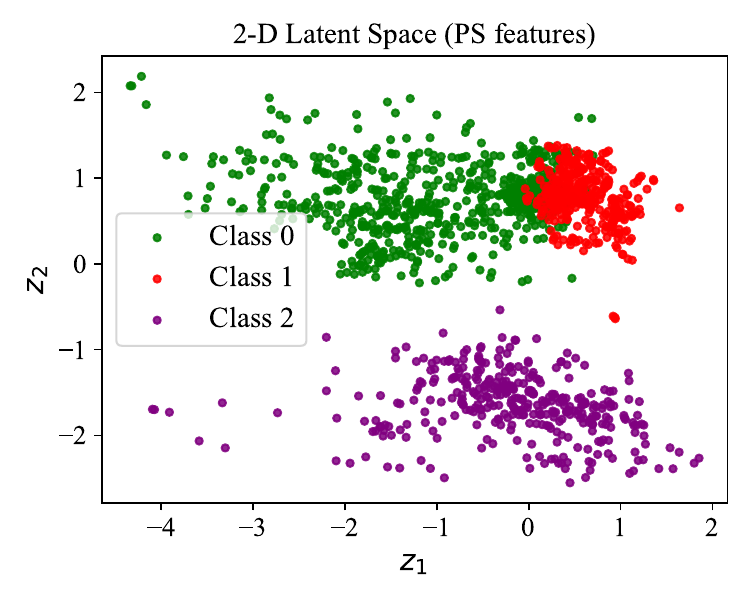}}
	\subfloat[5kHz, PS features\label{fig:latent-space-5_is}]{%
		\includegraphics[width=0.5\columnwidth]{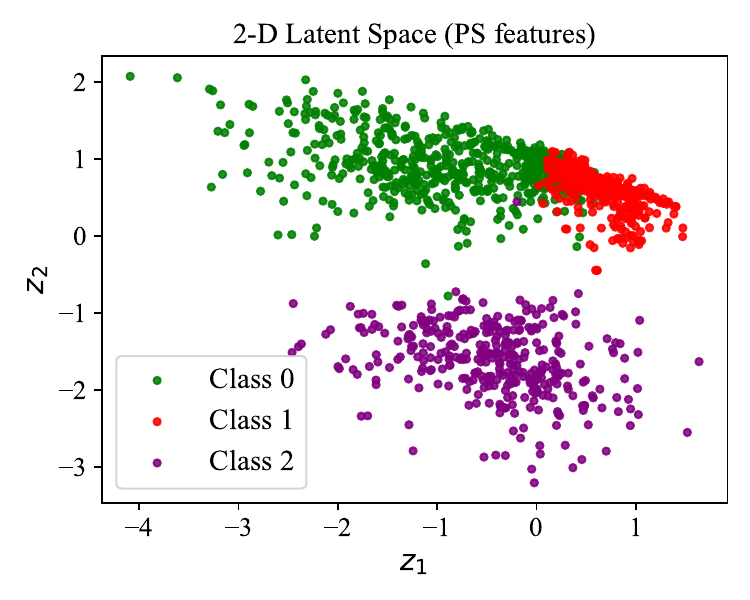}}
	\caption{Latent-space visualization on datasets under four sampling rates and two feature sets.}  
	\label{fig:latent-space} 
\end{figure*}

By default, features are extracted from both the F and PS stages, as described in Section~\ref{subsection-feature-extraction}. Considering that the F-stage is relatively short (in some cases less than 1 ms) and that the majority of measurements are sampled during the PS-stage, we also evaluate the localization performance using PS-stage features only. Finally, 8 models (4 sampling rates $\times$ 2 feature sets) were trained and compared.

Figs.~\ref{fig:ce-kl} illustrates the cross entropy (CE) loss and KL divergence recorded during training for 8 localization models. \ref{fig:ce} and \ref{fig:kl} were obtained using all features, whereas \ref{fig:ce_is} and \ref{fig:kl_is} were obtained using PS-stage features only. It is obvious that datasets with higher sampling rates converge to lower values in both metrics using both feature sets, showing superior predictive accuracy and compression (generalization) capability. This is because datasets with higher sampling rates inherently contain richer information about the GFs, while also having more redundant information that can be filtered out. Furthermore, an increase in KL divergence is observed in the early training phase, which is a normal phenomenon of VIB, as the encoder tends to pass more information to $\mathbf{z}$ initially to optimize its representation performance.

Fig.~\ref{fig:latent-space} visualizes the latent-space representations of the test sets for 8 localization models. \ref{fig:latent-space-100} to \ref{fig:latent-space-5} applied all features, whereas \ref{fig:latent-space-100_is} to \ref{fig:latent-space-5_is} applied PS-stage features. Under the full-feature setting, it is evident that as the sampling rate increases, both the density of clusters (of different labels) and the distance between clusters grow larger, thereby making the clusters easier to be separated by the decoder. Moreover, under a lower sampling rate, labels `0' and `1' exhibit overlap in certain regions, highlighting the challenge of distinguishing the GF with $N_{\rm sc}=0$. In contrast, label `2' (GF with $N_{\rm sc}>0$) is clearly separable because its fault characteristics are more pronounced. Compared with full-feature settings, using only PS-stage features results in looser clusters. Nevertheless, label `2' remains separable under all sampling rates, which indicates that PS-stage features are sufficiently informative to recognize GFs with $N_{\rm sc}>0$.

A detailed summary of performance is provided in Table~\ref{tab:eval_results}. Under the full-feature setting, when the sampling rate decreases from 100 kHz to 5 kHz, the overall accuracy falls from 99.9\% to 93.0\%. Notably, the recall for label `1' drops more sharply from 99.7\% down to 86.0\%, while recall for label `2' remains nearly constant at 99.4–100.0\%. When F-stage features are omitted, the overall accuracy declines further by up to 5\%. The recall for label `1' decreases significantly, whereas the recalls for labels `0' and `2' are only slightly affected. These results align with the latent-space observation and underscore that a higher sampling rate or the inclusion of F-stage features can improve GF localization especially for faults occurring at the negative terminal. GFs with $N_{\rm sc}>0$ are much easier to be recognized even under lower sampling rates and the omission of the F-stage features.

\begin{table}[htbp]
	\centering
	\caption{Evaluation results on different datasets.}
	\label{tab:eval_results}
	\begin{tabular}{c c c c c}
		\toprule
		\textbf{Dataset} & \textbf{Accuracy} & \textbf{Recall} (`0') & \textbf{Recall} (`1') & \textbf{Recall} (`2') \\
		\midrule
		100 kHz &  99.9\% & 100.0\% & 99.7\% & 100.0\% \\
		10 kHz &  97.7\% & 97.6\% & 95.5\% & 100.0\% \\
		7.5 kHz &  96.3\% & 96.2\% & 93.3\% & 99.4\% \\
		5 kHz &  93.0\% & 93.3\% & 86.0\% & 99.4\% \\
		100 kHz (PS) &  98.2\% & 98.3\% & 96.1\% & 100.0\% \\
		10 kHz (PS) &  93.0\% & 94.7\% & 82.7\% & 100.0\% \\
		7.5 kHz (PS) &  92.7\% & 92.3\% & 86.3\% & 99.7\% \\
		5 kHz (PS) &  91.6\% & 92.6\% & 81.3\% & 99.7\% \\
		\bottomrule
	\end{tabular}
\end{table}

\section{Conclusions}\label{section-conclusion}
This paper presented a string-level GF localization approach for 3$\phi$-TN PV systems based on edge-AI. We developed a PLECS-based simulation model incorporating a dynamic PV model that captures PV's hysteresis effects. Through fault-current analysis, we characterized three types of phase-fault-currents: lower-leg, upper-leg, and zero phase-fault-current. Moreover, a case study on different GF locations were presented. Building on these insights, we proposed an offline edge-AI development pipeline that generates diverse simulation cases under various fault scenarios, extracts correlation-based features from inverter's four-stage shutdown procedure, and trains a lightweight VIB-based localization model. Experiments on datasets with multiple sampling rates confirmed the strong performance of the proposed approach, achieving up to 99.9\% accuracy at 100~kHz and 93.0\% at 5~kHz with full features from F and PS stages. Faults with $N_{\rm sc}>0$ are highly distinguishable, whereas faults at the negative terminal pose greater challenges but can still be recognized effectively with increased sampling rate and the inclusion of F-stage features.



\bibliographystyle{IEEEtran}

\bibliography{reference}

\end{document}